\documentstyle[12pt,epsf]{article}
   \input{epsf}
\def\beq{\begin{equation}} \def\eeq{\end{equation}}
\def\lsim{\raise0.3ex\hbox{$<$\kern-0.75em\raise-1.1ex\hbox{$\sim$}}}
\def\gsim{\raise0.3ex\hbox{$>$\kern-0.75em\raise-1.1ex\hbox{$\sim$}}}
\def\H{{\scriptscriptstyle H}} 

\def\FPI{f_\pi}\def\fpi{\ifmmode\FPI\else$\FPI$\fi}
\def\MH{m_\H}\def\mh{\ifmmode\MH\else$\MH$\fi}
\def\LMH{\mu_\H}\def\lmh{\ifmmode\LMH\else$\LMH$\fi}
\def\FMH{\bar\mu_\H}\def\fmh{\ifmmode\FMH\else$\FMH$\fi}
\def\PPB{\langle\bar\psi\psi\rangle}\def\ppb{\ifmmode\PPB\else$\PPB$\fi}
\def\NT{N_\tau}\def\t{\ifmmode\NT\else$\NT$\fi}
\def\NZ{N_z}\def\z{\ifmmode\NZ\else$\NZ$\fi}
\def\TC{T_c}\def\tc{\ifmmode\TC\else$\TC$\fi}
\def\PQ{\langle P\rangle}\def\pq{\ifmmode\PQ\else$\PQ$\fi}
\def\PL{{\sl Phys.\ Lett.\/}~}        \def\PR{{\sl Phys.\ Rev.\/}~}
\def\PRL{{\sl Phys.\ Rev.\ Lett.\/}~} \def\NP{{\sl Nucl.\ Phys.\/}~}

\begin{document}
{\thispagestyle{empty}
\noindent \hspace{1cm}  \hfill November 1994 \hspace{1cm}\\
\mbox{}                 \hfill BI-TP 94/55    \hspace{1cm}\\

\begin{center}\vspace*{1.0cm}
{\large\bf Three-Dimensional SU(3) gauge theory and the}\\
\vspace*{0.2cm}
{\large\bf Spatial String Tension of the}\\
\vspace*{0.2cm}
{\large\bf (3+1)-Dimensional Finite Temperature SU(3) Gauge Theory}
\\\vspace*{1.0cm}
{\large F.~Karsch, E. Laermann and M. L\"utgemeier}\\
\vspace*{1.0cm}{\normalsize
{Fakult\"at f\"ur Physik \\
Universit\"at Bielefeld \\
P.O.Box 100131 \\
D-33501 Bielefeld\\
Germany.}}\\
\vspace*{2cm}{\large \bf Abstract}
\end{center}

\setlength{\baselineskip}{1.3\baselineskip}

We establish a close relation between the spatial string tension of
the (3+1)-dimensional $SU(3)$ gauge theory at finite temperature
($\sigma_s$) and the string tension of the 3-dimensional $SU(3)$ gauge
theory ($\sigma_3$) which is similar to what has been found previously
for $SU(2)$. We obtain $\sqrt{\sigma_3} = (0.554 \pm 0.004) g_3^2$
and $\sqrt{\sigma_s} = (0.586 \pm 0.045)g^2(T) T$, respectively. For
temperatures larger than twice the critical temperature results are
consistent with a temperature dependent coupling running according to
the two-loop $\beta$-function with $\Lambda_T = 0.118(36)T_c$.
}
\newpage
\setcounter{page}{1}
\setlength{\baselineskip}{1.5\baselineskip}
\section{INTRODUCTION}

$SU(N)$ gauge theories at high temperature are known to undergo a
phase transition to a deconfined phase. At the phase transition point
basic thermodynamic observables show a drastic qualitative change,
which suggests that the high temperature phase consists of weakly
interacting, {\it asymptotically free} partons.
The heavy quark potential, for instance, changes from a confining
potential in the low temperature phase to a Debye screened potential at
high temperature. At the same time there exist, however, correlation
functions, whose structure
does not change qualitatively at $T_c$, although there may be
significant changes in their temperature dependence.
For instance, the pseudo-potential, extracted
from space-like Wilson loops, is known to be confining for all
temperatures \cite{Borgs,Polonyi}. This has been taken
as indication for the survival of certain {\it confining} properties in the
high temperature phase.
The different high temperature behaviour of observables like the heavy
quark potential and the spatial string tension can be
understood at least qualitatively in terms of the structure of the
effective, three-dimensional theory which describes the high temperature
phase of QCD. This effective theory,
obtained from {\it dimensional reduction} \cite{reisz},
is a 3-$d$ gauge-Higgs model with
Higgs-fields in the adjoint representation. The gauge sector of the
model is confining and
at the same time it leads to a screened Coulomb potential for the
adjoint Higgs fields, which represent the static 0-components of the
gauge potential in (3+1) dimensions.

It recently has been shown \cite{Bali,Leo} that in the case of an $SU(2)$
gauge theory
the resulting spatial string tension remains temperature independent up
to $T_c$ and then starts rising rapidly. A similar behaviour has been
found in lower dimensions and also in $Z(2)$ gauge theories
\cite{CaselleA,CaselleB}. The qualitative behaviour of the spatial
string tension thus seems to be a generic feature of the finite
temperature phase transition in confining gauge theories.
The temperature dependence may be qualitatively understood in terms of
string fluctuations, which get suppressed due to the reduction of one
space-time dimension with increasing temperature. This leads to a more
rigid string, i.e. a larger string tension \cite{CaselleB}.

In a recent high statistics Monte Carlo study of the $SU(2)$ gauge
theory \cite{Bali} it could be shown that the linear slope of the
pseudo-potential extracted from spatial Wilson loops - the {\it spatial
string tension} $\sigma_s$ - is closely related to the string tension of 
the gauge sector of the 3-$d$ effective theory, $\sigma_3$, alone. 
The Higgs part does not seem
to contribute substantially to this quantity. In fact, for
temperatures larger than twice $T_c$ the spatial string tension has been
found to rise like $g^4(T) T^2$ with a proportionality constant, which
is only 10\% larger than that for the string tension in a 3-$d$ gauge
theory. The running
coupling, however, turned out to be quite large, $g^2(2T_c) \simeq 2.7$.

It is the purpose of the present paper to study the relation between
$\sigma_s$ and $\sigma_3$ in the case of the SU(3) gauge theory in (3+1)
and 3 dimensions, respectively. Unlike in the case of $SU(2)$, where the
string tension in 3-$d$ had been studied quite accurately \cite{Teper},
we have to start here with a determination of $\sigma_3$ for the
3-$d~SU(3)$ gauge theory. We will discuss this calculation in section 2.
In section 3 we describe the calculation of the spatial string tension of
the finite temperature $SU(3)$ gauge theory in (3+1) dimensions for three
values of the temperature. Both results will be discussed in section 4.

\section{STRING TENSION OF THE 3-D SU(3) GAUGE THEORY}

We have analyzed the heavy quark potential in a 3-$d~SU(3)$ gauge
theory on lattices of size $32^3$. The 3-$d$ Wilson action is given by
\beq
S_3 = \beta_3 \sum_{0 \le \mu < \nu \le 2} (1 - {1 \over 3} Re~{\rm Tr}
U_{x,\mu} U_{x + \mu, \nu} U^\dagger_{x+\nu,\mu} U^\dagger_{x, \nu})~~~.
\label{action}
\eeq
Here $\beta_3 = 6/ag_3^2$ denotes the dimensionless coupling, which in the
continuum limit is related to the lattice spacing ``a'' and the
dimensionful gauge coupling, $g^2_3$. The latter sets the scale for all
dimensionful quantities in this limit. Lattice observables thus will
scale with appropriate powers of the coupling $\beta_3$.

The heavy quark potential has been calculated from ratios of smeared
Wilson loops, $W(R,S)$, \cite{Bali} as
\beq
V(R) a = - \lim_{S \to \infty} \ln {{W(R, S+1)} \over {W(R,S)}}~~~.
\label{potential}
\eeq
We find that generally $S > S_{min} \simeq 4$
is sufficient to reach a plateau in the ratio of smeared Wilson
loops. We then extract $V(R) a$ from a fit to ratios with
$S > S_{min}$.

The potential has been calculated at six values of $\beta_3$ in
the interval $18 \le \beta_3 \le 24$.
The results are shown in Fig. 1. At each value of the coupling
we have performed 37.500 overrelaxed/heat-bath
updates and have analyzed 1500 configurations separated 
by 5 iterations (1 iter. $\equiv$ 4OR+1HB). Errors have been obtained 
from a jackknife-analysis.

\begin{figure}[htb]\vskip85mm
\includegraphics{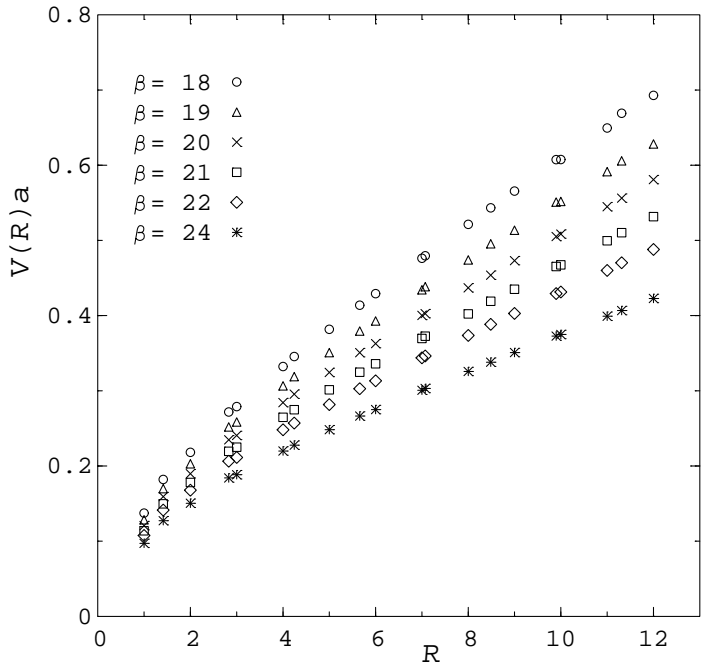}
\caption{The heavy quark potential in 3-$d~SU(3)$ gauge theory obtained
from calculations on a $32^3$ lattice. The errors are smaller than the size
of the symbols.}
\label{VvsR}
\end{figure}

At short distances the 3-$d$ potentials clearly differ from their 4-$d$
counterparts. They show a much weaker $R$-dependence\footnote{Through out 
this paper $R$ is given in lattice units.} in this regime 
as may be expected from a logarithmic Coulomb potential.
The long-distance part of the heavy quark potential is dominated by
the linearly rising confinement part, $V_{conf} \sim \hat\sigma_3 R$.
In addition, however, one expects to find also a universal
contribution from string fluctuations, ${\pi (d-2) /24 R}$, which
might be visible at intermediate distances \cite{Alvarez}.

We extract the string tension from two and three parameter fits to the
long distance part of the potential using the general ansatz

\beq
V(R) a = v - {{c} \over R} + \hat\sigma_3 R~~~{\rm for}~~
R > R_{min}~~.
\label{fitI}
\eeq
In the case of two parameter fits we fix the coefficient of the $1/R$-term 
to the value of the string model, $c=\pi/24$, and vary $R_{min}$
until a plateau is found for the coefficient of the linear term,
$\hat\sigma_3(R_{min})$. We find that this coefficient rises with increasing
$R_{min}$. We need $R_{min} \simeq 3$ for $\beta_3 = 18$ and $R_{min} = 6$ 
for $\beta_3 = 24$ until a plateau is reached within statistical errors.
In the case of three parameter fits we proceed in the same way. Here we
find stable string tension values for $R_{min} \ge 4$. We have 
verified that the results of both types of fits agree within statistical
errors \cite{lattice}. 
Our final results for $\hat\sigma_3$, obtained from two parameter fits as
described above, are summarized in Table 1. 
\begin{table}
\begin{center}
  \arraycolsep5mm
  \renewcommand{\arraystretch}{1.50}
  \[ 
  \begin{array}[t]{|@{\quad}c|c@{\quad}|}
    \hline
    \beta_3 & \hat{\sigma}_3 \\
    \hline
    18.0 & 0.042895 \;\, (97) \\
    19.0 & 0.037818 \;\, (94) \\
    20.0 & 0.034100 \;\, (90) \\
    21.0 & 0.030716 \;\, (78) \\
    22.0 & 0.027406 \;\, (75) \\
    24.0 & 0.022835 \;\, (71) \\
    \hline
  \end{array}
  \] 
\end{center} 
\caption{The three-dimensional string tension calculated on lattices of size
$32^3$ at various values of $\beta_3$}
\label{TableI}
\end{table}
In the continuum limit the string tension scales like
\beq
\sqrt{\hat\sigma_3} = {b_1 \over {\beta_3}} +
{b_2 \over{\beta_3^2}}~~~.
\label{sigma}
\eeq
The results for $\hat\sigma_3 \beta_3$ obtained from our two parameter
fits are shown in Fig.~2. These data can be fitted
to the form given by eq.~(4). We find,
\beq
\sqrt{\hat\sigma_3}\beta_3 = 
3.326(22) + 7.2(4)/\beta_3~~,
\label{sbfit}
\eeq
which gives for the string tension in the continuum limit, $\sigma_3 a^2
\equiv \hat\sigma_3$, 
\beq
\sqrt{\sigma_3} = (0.554 \pm 0.004) g^2_3~~~.
\label{sqrtsigma}
\eeq

\begin{figure}[htb]\vskip85mm
\includegraphics{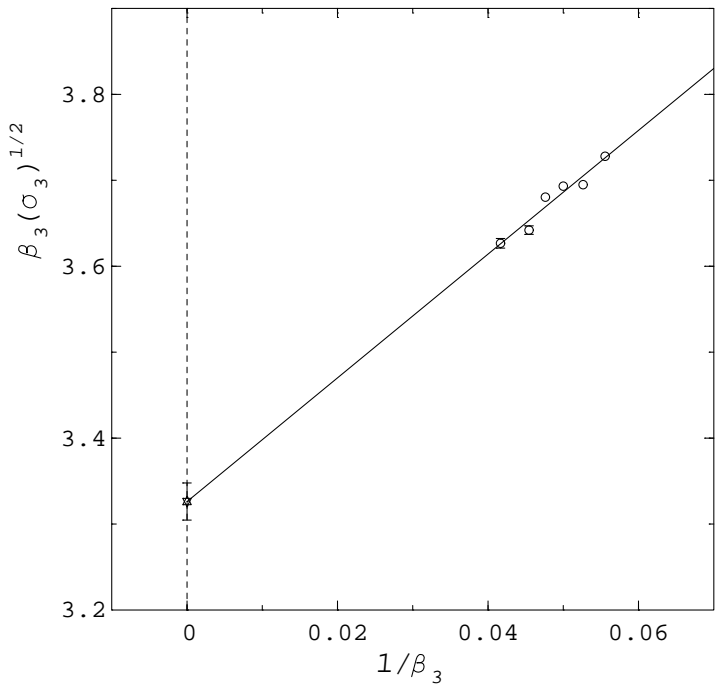}
\caption{Scaling behaviour of the string tension in 3-$d~SU(3)$ gauge
theory. Errors on the data points are of the size of the symbols.
The solid line is the fit given by Eq. 5, the star at $1/\beta_3 = 0$ 
shows the constant appearing in this fit.}
\label{betasigma}
\end{figure}

\section{THE SPATIAL STRING TENSION}

Let us now come to a discussion of the spatial string tension of the
$SU(3)$ gauge theory at finite temperature. In analogy to the
analysis performed in the $SU(2)$ case \cite{Bali} we study the spatial
pseudo-potentials resulting from Wilson loops in hyperplanes orthogonal
to the temporal direction of the lattice. At short distances these
pseudo-potentials have been studied previously and have been shown to agree
quite well with results obtained within the dimensionally reduced effective 
theory at high temperatures \cite{karkkainen}. Here we aim at an analysis
of the large distance behaviour of these pseudo-potentials using smeared
Wilson loops as discussed in the previous section. 
Our calculations have been
performed at a fixed value of the gauge coupling,
$\beta = {6/g^2} = 6.0$, i.e.~at fixed lattice cut-off.
The temperature is varied by varying the temporal extent, $N_\tau$,
of a lattice of size $32^3 \times N_\tau$.
The deconfinement
transition on a lattice with temporal extent $N_\tau = 8$ is known
to occur close to this value of the gauge coupling \cite{Columbia}.
For the purpose of the present calculation it is sufficient to identify
$\beta=6.0$ with the critical coupling on lattices with temporal extent
$N_\tau=8$. We have performed simulations for $N_\tau =2$, 3 and 4.
These values thus
cover the temperature interval, $2 \lsim {T/ T_c}\lsim 4$. On these
lattices we have collected between 200 and 400 configurations
separated by 50 iterations (1 iter. $\equiv$ 7OR+1HB).
\par
The spatial pseudo-potentials, extracted from smeared spatial Wilson loops 
are shown in Fig. 3. In (3+1) dimensions at
zero temperature the Coulomb term in the heavy quark potential is a
$1/R$-term. This is expected to transform into a logarithmic behaviour at very 
high temperatures. In order to avoid a prejudice on the subleading $R$ dependence
at large distances, we have extracted the string tension from a 
linear fit
\beq
V(R) a = v + \hat\sigma_s R~~~{\rm for}~~
R > R_{min}~~,
\label{fitII}
\eeq 
where we have varied the lower limit of the fitting range, $R_{min}$, until
the string tension has been found to be independent of it within errors.
For all three temperatures we find this to be the case for $R_{min} > 4$
\cite{lattice}.
The resulting values for the spatial string tension are summarized
in Table 2.
\begin{table}
\begin{center}
  \arraycolsep5mm
  \renewcommand{\arraystretch}{1.50}
  \[ 
  \begin{array}[t]{|c|c|c|}
    \hline
    N_{\tau} & T/T_c & \hat{\sigma}_s \\
    \hline
    32 &  0.25 & 0.0513 \;\pm\, 0.0025 \\
    4 &  2   & 0.0874 \;\pm\, 0.0014 \\
    3 & 2.67 & 0.1224 \;\pm\, 0.0052 \\
    2 &  4   & 0.2384 \;\pm\, 0.0090 \\
    \hline
  \end{array}
  \]
\end{center}
\caption{The spatial string tension calculated on lattices of size
$32^3\times N_\tau$ for $N_\tau = 2$, 3, 4 (this paper) and 32 
([14]) at $\beta = 6.0$}
\label{TableII}
\end{table}

The detailed investigation performed in the case of $SU(2)$ has shown that
the temperature dependence of the spatial string tension above $2T_c$ is
consistent with what one would expect from dimensional reduction if the
Higgs-sector does not significantly contribute to the
string tension, $\sqrt{\sigma_s} \sim g^2(T) T$, with $g^2(T)$ given by
the 2-loop $\beta$-function. The same does seem to hold true in the case
of $SU(3)$ as can be seen from Fig.~4 where we show the ratio
$T/\sqrt{\sigma_s}$ from our present analysis of the $SU(3)$ gauge theory
as well as the data for $SU(2)$ \cite{Bali}. Also shown in this figure
is $T_c/\sqrt{\sigma}$, where $\sigma$ is the zero temperature ($T/T_c
\simeq 0.25$) string tension obtained on a $32^4$ lattice
\cite{Schilling}, and a first result from an ongoing more detailed
investigation of the spatial string tension on lattices with temporal
extent $N_\tau =6$ ($T/T_c \simeq 1.33$) \cite{quadrics}. In both cases 
the string tension has also been calculated at $\beta=6.0$.

\begin{figure}[htb]\vskip85mm
\includegraphics{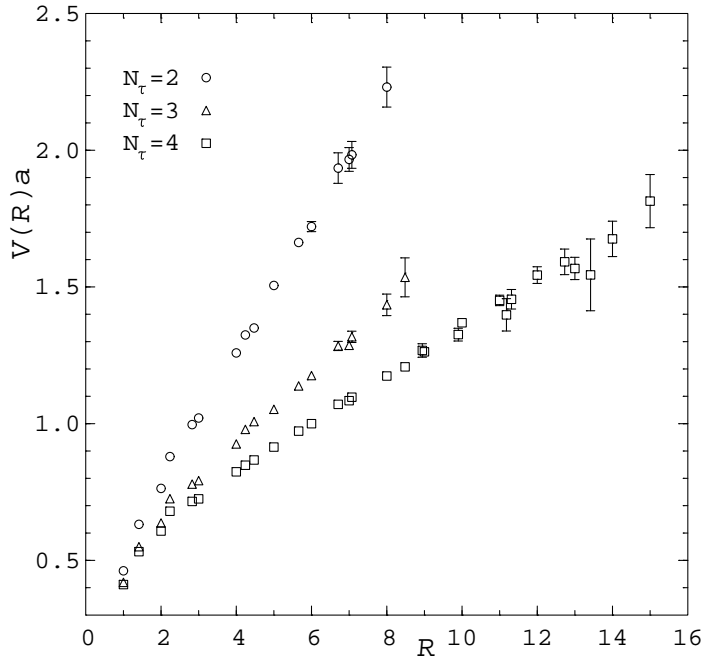}
\caption{The heavy quark potential in (3+1)-dimensional $SU(3)$ gauge theory 
obtained from calculations on  $32^3\times N_\tau$ lattices, 
with $N_\tau =2$ (circles), 3 (triangles) and 4 (squares) at $\beta=6.0$.}
\label{Tpotential}
\end{figure}

Assuming the validity of 2-loop scaling for the running coupling constant,
\beq
g^{-2} (T) = {11 \over 8 \pi^2}\ln\bigl(T/\Lambda_T \bigr) +
{51 \over 88 \pi^2} \ln\bigl(2\ln(T/\Lambda_T)\bigr) ~~,
\label{running}
\eeq
for temperatures larger than twice $T_c$ we
find  for the spatial string tension of (3+1)-dimensional $SU(3)$,
\beq
\sqrt{\sigma_s(T)}\;=\; (0.586 \pm 0.055) g^2(T) T \quad,\quad T\gsim 2T_c~~.
\label{spatialfit}
\eeq
Here the additional free parameter in the
$\beta$-function is determined as $\Lambda_T = 0.118(36)T_c$.

\begin{figure}[htb]\vskip85mm
\includegraphics{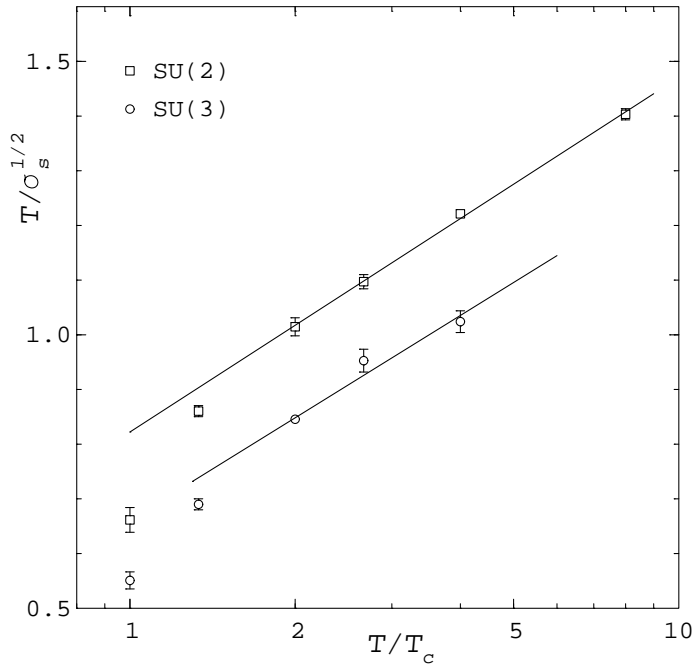}
\caption{Temperature over square root of the spatial string tension
versus $T/T_c$. Shown is a comparison of results for the $SU(2)$ 
and $SU(3)$ gauge groups.}
\label{Toversigma}
\end{figure}

\section{DISCUSSION}

We have studied the string tension in three-dimensional $SU(3)$ gauge
theory and performed a first
exploratory investigation of the spatial string tension in (3+1)-dimensional
$SU(3)$ at finite temperature. We find a behaviour which closely resembles
results found in the case of an $SU(2)$ gauge theory: For temperatures above
$2T_c$ the spatial string tension is well described by the string tension
of a three-dimensional gauge theory, $\sqrt{\sigma_s} \simeq \sqrt{\sigma_3}$,
with $g_3^2 \equiv g^2(T)T$.
This gives further support to the observation that the Higgs-sector in
dimensionally reduced QCD does not contribute significantly to the spatial
string tension.

From the temperature dependence of $\sqrt{\sigma_s}$ above $2T_c$ we find
evidence for a running coupling constant. The coupling $g^2(T)$ turns out to
be somewhat smaller than in the case of $SU(2)$ at comparable temperatures.
For instance we find $g^2(2T_c) \simeq 2.0$ for the $SU(3)$ gauge theory
compared to 2.7 in the case of $SU(2)$. Similarly the scale parameter in the
perturbative 2-loop $\beta$-function is slightly larger. It is, however, 
interesting to note that in both cases the value of $\Lambda_T$ turns out to 
be of the order of 10\% of the scale $\Lambda_{\overline{MS}}$ determined
in pure gauge (zero flavour) simulations at zero temperature for instance 
from the heavy quark potential \cite{Schilling}.
In fact, it is consistent with the relation, $\Lambda_T =
\Lambda_{\overline{MS}} /4\pi$, a combination which naturally occurs in the
renormalization scale dependent terms in higher order perturbative
calculations\footnote{In Ref.\cite{Farakos} the 
relation, $\Lambda_T =\Lambda_{\overline{MS}} /4\pi\exp\{-\gamma_E\}$, has 
been suggested for the $\Lambda$-parameter entering the running coupling in 
the effective 3-$d$ theory obtained by dimensional reduction.}.
\cite{Farakos,Arnold}
\vskip 20pt
\noindent
{\large \bf Acknowledgements:}

\noindent
We thank J. Fingberg and U. Heller for helpful discussions.
The computations have been performed on the Cray Y-MP at HLRZ and the
Q1 Quadrics at the University of Bielefeld. The work has been supported
in part by the Deutsche Forschungsgemeinschaft under contracts Pe
340/3-2 and 340/6-1.


\begin{thebibliography}{99}
\bibitem{Borgs}
C. Borgs, \NP B261 (1985) 455.
\bibitem{Polonyi}
E. Manousakis and J. Polonyi, \PRL 58 (1987) 847.
\bibitem{reisz}
T. Reisz, Z. Phys. C - Particles and Fields 53 (1992) 169 and references
therein.
\bibitem{Bali}
G.S. Bali, J. Fingberg, U.M. Heller, F. Karsch and K. Schilling,
\PRL 71 (1993) 3059.
\bibitem{Leo}
L. K\"arkk\"ainen, P. Lacock, D.E. Miller, B. Petersson and T. Reisz,
\PL B312 (1993) 173.
\bibitem{CaselleA}
M. Caselle, R. Fiore, F. Gliozzi, P. Guaita and S. Vinti,
\NP B422 (1994) 397.
\bibitem{CaselleB}
M. Caselle and A. D'Adda, {\it The Spatial String Tension in High
Temperature Lattice Gauge Theories}, DFTT 8/94, March 1994.
\bibitem{Teper}
M. Teper, \PL B311 (1993) 223.
\bibitem{karkkainen}
L. K\"arkk\"ainen, P. Lacock, D.E. Miller, B. Petersson and T. Reisz,
\PL B282 (1992) 121.
\bibitem{Columbia}
S.A. Gottlieb, J. Kuti, D. Toussaint, A.D. Kennedy, S. Meyer,
B.J. Pendleton and R.L. Sugar, \PRL 55 (1985) 1958; \\
H. Ding and N. Christ, \PRL 60 (1988) 1367.
\bibitem{Alvarez}
M. L\"uscher, K. Symanzik and P. Weisz, \NP B173 (1980) 365; \\
O. Alvarez, \PR D24 (1981) 440.
\bibitem{lattice}
M. L\"utgemeier, contribution to {\it Lattice 94}, to appear in \NP
B (Proc.Suppl.).
\bibitem{Schilling}
G.S. Bali and K. Schilling, \PR D47 (1993) 661.
\bibitem{quadrics}
G. Boyd, J. Engels, F. Karsch, E. Laermann, C. Legeland,
M. L\"utgemeier and B. Petersson, {\it work in progress}.
\bibitem{Farakos}
K. Farakos, K. Kajantie, K. Rummukainen and M. Shaposhnikov, \NP B425 
(1994) 67.
\bibitem{Arnold}
P. Arnold and C. Zhai, {\it The three-loop free energy for pure gauge QCD},
University of Washington preprint, UW/PT-94-03.
\end{thebibliography}
\end{document}